\documentclass[10pt,twocolumn]{article}
\usepackage{times}  %
\usepackage{helvet}  %
\usepackage{courier}  %
\usepackage[hyphens]{url}  %
\usepackage{graphicx} %
\usepackage[numbers]{natbib} 
\usepackage{caption} %
\DeclareCaptionStyle{ruled}{labelfont=normalfont,labelsep=colon,strut=off} %
\usepackage{algorithm}
\usepackage{algorithmic}

\usepackage{newfloat}
\usepackage{listings}
\floatstyle{ruled}
\newfloat{listing}{tb}{lst}{}
\floatname{listing}{Listing}

\usepackage{comment}
\usepackage{caption}
\usepackage{subcaption}
\usepackage{color}
\usepackage{xspace}

\newcommand{\one}{({\em i}\/)\xspace}
\newcommand{\two}{({\em ii}\/)\xspace}
\newcommand{\three}{({\em iii}\/)\xspace}

\def\eg{\emph{e.g.}\xspace}
\def\ie{\emph{i.e.}\xspace}

\newcommand{\pb}[1]{\vspace{0.75ex}\noindent{\bf \em #1}\hspace*{.3em}}

\date{\vspace{-3ex}}

\title{Flocking to Mastodon: Tracking the Great Twitter Migration}
\author {
    Haris Bin Zia,\textsuperscript{\rm 1}
    Jiahui He,\textsuperscript{\rm 2} 
    Aravindh Raman,\textsuperscript{\rm 3}
    Ignacio Castro,\textsuperscript{\rm 1}
    Nishanth Sastry,\textsuperscript{\rm 4}
    Gareth Tyson\textsuperscript{\rm 2,1}\\[0.25ex]
    \small 
    \textsuperscript{\rm 1}Queen Mary University of London 
    \textsuperscript{\rm 2}Hong Kong University of Science and Technology (GZ)\\\small 
    \textsuperscript{\rm 3}Telefonica Research 
    \textsuperscript{\rm 4}University of Surrey\\
}

\graphicspath{ {./images/} }

\begin{document}

\maketitle

\begin{abstract}

The acquisition of Twitter by Elon Musk has spurred controversy and uncertainty among Twitter users. The move raised as many praises as concerns, particularly regarding Musk's views on free speech. As a result, a large number of Twitter users have looked for alternatives to Twitter. Mastodon, a decentralized micro-blogging social network, has attracted the attention of many users and the general media. In this paper, we track and analyze the migration of 136,009 users from Twitter to  Mastodon. Our analysis sheds light on the user-driven pressure towards centralization in a decentralized ecosystem and identifies the strong influence of the social network in platform migration. We also characterize the activity of migrated users on both Twitter and Mastodon.
\end{abstract}

\section{Introduction}
\label{sec:introduction}

In October 2022, Elon Musk, a
self-declared ``free speech absolutist'' acquired Twitter --- the social network that he regarded as the ``de facto town square'' where public debate takes place.
Musk's takeover has been controversial and highly publicized.
Some users admire Musk and his takeover, regarding it as crucial for free speech; others have expressed concerns over increased misinformation and toxicity. 

\begin{figure}[t]
     \centering
     
     \begin{subfigure}{\columnwidth}
         \centering
         \includegraphics[width=0.8\columnwidth]{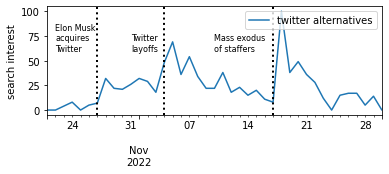}
         \caption{}
         \label{fig:gtrends1}
     \end{subfigure}

     \begin{subfigure}{\columnwidth}
         \centering
         \includegraphics[width=0.8\columnwidth]{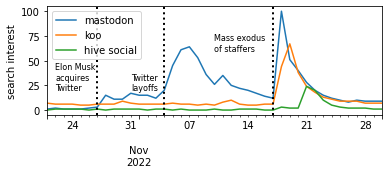}
         \caption{}
         \label{fig:gtrends2}
     \end{subfigure}
     
        \caption{Interest over time for the search terms (a) Twitter alternatives and (b) Mastodon, Koo \& Hive Social.}
        \label{fig:gtrends}
\end{figure}

Regardless of one's stance, it is undeniable that the acquisition has led to a series of noteworthy events.
On November 04, 2022, Musk fired half of the 7,500 employees previously working at Twitter. 
Two weeks later (November 17, 2022), 
hundreds of employees resigned in response to an
 ultimatum to commit to ``extremely hardcore'' work or leave.  
These events and the associated public backlash, prompted many users to search for alternatives. 
Figure~\ref{fig:gtrends1} presents a time series of Google trend search interest for ``Twitter alternatives''.
We observe a large spike on October 28, 2022, the day after Musk's takeover.
Similarly, Figure~\ref{fig:gtrends2} shows equivalent search interest for other popular alternatives to Twitter, \eg Koo (an Indian micro-blogging and social networking service), and Hive (a micro-blogging service that permits NSFW mature content). 

One platform that stands out as being particularly prominent is \emph{Mastodon}, a decentralized micro-blogging platform. Although released in 2016, Mastodon has anecdotally gathered significant attention since October 2022. It is part of the wider \emph{fediverse}, in which any person can create and operate their own Mastodon server (aka ``instance''). 
Each Mastodon instance operates as an independent microblogging service, where users can create local accounts and enjoy similar functions to Twitter (\eg posting, following). 
Importantly, these instances can also federate together, allowing users on one instance to follow users on another.
This means that Mastodon operates in a decentralized fashion (with people joining independent instances), while retaining the ability to interact across the entire globe.
This new paradigm has attracted significant attention and has made it an obvious candidate for users who are unhappy with the Musk acquisition (and the associated centralization of power in the hands of one individual).

This sudden interest in Mastodon offers a unique opportunity to study the migration of users between social networks. 
This is particularly the case due to the differing value propositions of the two platforms, with clear contrasts in the governance and ownership of Twitter vs. Mastodon. The unusual circumstances of the migration create further dimensions of analysis.
With this context in mind, we explore the following three research questions: 

\begin{itemize}
  \item \textbf{RQ1} How are new users spread across Mastodon instances, and are there any consequences for decentralization?
  \item \textbf{RQ2} How much (if at all) does a user's ego-centric Twitter follower network influence their migration to Mastodon?
  \item \textbf{RQ3} What are usage patterns of migrated users across both platforms?
\end{itemize}

To address these questions, we track 136,009 unique twitter users who moved to 2,879 unique Mastodon instances. %
The main findings related to our three RQs are as follows:

\begin{itemize}

  \item There is a user-driven pressure towards centralization on Mastodon (the top 25\% most populous instances contain 96\% of the users). This pressure is counterbalanced by the greater activity of the users on smaller instances. On average, users of single-user instances post 121\% more statuses than users on bigger instances.

  \item The social network of users on Twitter influences their choice of an instance on Mastodon \eg 4.09\% of users changed the instance on which they created an account (when they first migrated to Mastodon) and moved to the instance of choice of their Twitter followees who migrated to Mastodon as well. %

  \item Users tend to post different content across the two platforms. On average, only 1.53\% of a user's Mastodon posts are identical to their Twitter posts. Twitter hosts more diverse topics ranging from Entertainment to Politics, whereas discussions around Fediverse and Migration dominate on Mastodon.

\end{itemize}

\section{Mastodon Primer}
\label{sec:primerMast}

Mastodon is an open-source~\cite{mastodonGithub} federated server platform released in 2016. It offers micro-blogging functionality, allowing administrators to create their own independent Mastodon servers, aka \textbf{instances}. 
Each unique Mastodon instance works much like Twitter, allowing users to register new accounts and share statuses with their followers -- equivalent to tweeting on Twitter. Users can also \textbf{boost} others' statuses -- equivalent to retweeting on Twitter.

Instances can work in isolation, only allowing locally registered users to follow each other. 
However, Mastodon instances can also \textbf{federate}, whereby users registered on one instance can follow users registered on another instance.
This results in the instance \textbf{subscribing} to posts performed on the remote instance, such that they can be pushed across and presented to local users.
For simplicity, we refer to users registered on the same instance as \textbf{local}, and users registered on different instances as \textbf{remote}.
Note that a user registered on their local instance does \emph{not} need to register with the remote instance to follow the remote user. Instead, a user just creates a single account with their local instance; when the user wants to follow a user on a  remote instance, the user's local instance performs the subscription on the user's behalf. 
This process is implemented using an underlying subscription protocol, ActivityPub~\cite{activitypub}. This makes Mastodon compatible with other decentralised micro-blogging implementations (notably, Pleroma). 
The \textbf{Fediverse}, refers to the growing group of ActivityPub compatible, and therefore interconnected, applications. 

When a user logs in to their local instance, they are presented with three timelines: \one~a \textit{home} timeline, with statuses shared by the accounts whom the user follows; \two~a \textit{local} timeline, listing the statuses generated within the same instance; and \three~a \textit{federated} timeline, with \emph{all} statuses that have been retrieved from remote instances. 
The latter is not limited to remote statuses that the user follows; rather, it is the union of remote statuses retrieved by all users on the instance.

\section{Data Collection}
\label{sec:datacollection}

\subsection{Mastodon Accounts from Twitter}
\label{sec:mastodonaccountsfromtwitter}

\begin{figure}[t]
     \centering
     \includegraphics[width=0.8\columnwidth]{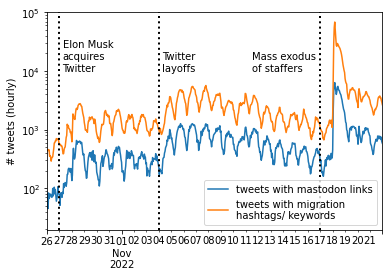}
     \caption{Temporal distribution of tweets containing (i) links to Mastodon instances and (ii) migration related keywords/ hashtags.}
     \label{fig:datacrawled}
\end{figure}

We collect a global list of Mastodon instances from \url{instances.social}, which  contains a comprehensive index of Mastodon instances. 
We compile a set of 15,886 unique instances.
We then collect all tweets containing a link to these Mastodon instances using Twitter’s Search API.\footnote{https://api.twitter.com/2/tweets/search/all} 
Additionally, we collect all tweets containing the following list of keywords related to the migration from Twitter: `mastodon', `bye bye twitter', `good bye twitter'; and hashtags \#Mastodon, \#MastodonMigration, \#ByeByeTwitter, \#GoodByeTwitter, \#TwitterMigration, \#MastodonSocial, \#RIPTwitter.
In total, we collect 2,090,940 tweets posted by 1,024,577 users between October 26, 2022 (\ie a day before Musk's takeover) and November 21, 2022. Figure~\ref{fig:datacrawled} shows the temporal distribution of these tweets.  

We next search for Mastodon usernames in these tweets and the accompanying metadata of any account that posted a tweet (\ie display name, location, description, URLs, pinned tweet text).
Mastodon usernames take the form @alice@example.com and https://example.com/@alice, where alice is a username and \url{example.com} is an instance.
To map a Mastodon handle to a Twitter account, we do this search in a hierarchical fashion: We first look for Mastodon usernames in user metadata (\eg bio) and create a mapping between Twitter account \& Mastodon account if one is found. If the search is unsuccessful at the first step, we then look for Mastodon usernames in the tweet text. 
To ensure mapping accuracy, we only map a Twitter account to a Mastodon account identified from a tweet text if both the Twitter and Mastodon usernames are identical.

Using this methodology, we identify the Mastodon accounts of 136,009 Twitter users, which are created across 2,879 unique Mastodon instances. %
We find that 72\% of Twitter users that migrated created a Mastodon account with the same username that they use on Twitter. 
4\% of the Twitter users who create a Mastodon account, have a (legacy) verified status (\ie authentic, notable, and active) on Twitter, suggesting that even well-established users have been migrating. %

\begin{figure}[t]
     \centering
     \includegraphics[width=0.8\columnwidth]{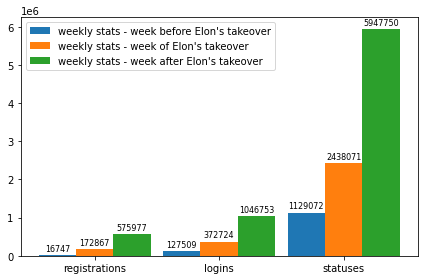}
     \caption{Weekly activity on Mastodon instances.}
     \label{fig:mastodonactivity}
\end{figure}

While we track and analyze the migration of a large number of Twitter users (136,009), the takeover of Twitter by Musk, is likely to have pushed even more users to migrate who the above methodology cannot identify. Indeed, on November 12, 2022, Mastodon announced that over 1 million users had registered since October 27, 2022~\cite{mastodontweet}, significantly more than our methodology identifies.
To understand the wider activities on the Mastodon instances, we cross-check the new registrations on the 2,879 instances by crawling their weekly activity from Mastodon's Weekly Activity Endpoint.\footnote{https://docs.joinmastodon.org/methods/instance/\#activity} 
Figure~\ref{fig:mastodonactivity} shows the weekly number of registrations, logins and statuses. We notice a large increase in all three activity metrics after the Twitter acquisition.
Of course, we cannot confirm that all these users migrated directly from Twitter. However, given the timeline of registrations, we believe that it is very likely that a large share of these new users migrated from Twitter.

\subsection{Twitter and Mastodon Timelines.}
\label{sec:twitterandmastodontimelines}

We next crawl both the Twitter and Mastodon timelines of the migrating users identified in the previous section. 
We use Twitter's Search API and Mastodon's Account's Statuses Endpoint.\footnote{\url{https://docs.joinmastodon.org/methods/accounts/#statuses}}
For each user, we crawl all tweets/statuses from October 01, 2022 to November 30, 2022. In total, we gather Twitter timelines for 94.88\% of the users.
The rest were suspended (0.08\%), deleted/deactivated (2.26\%), or the tweets were protected (2.78\%). 
We crawl the timelines of 79.22\% of Mastodon users: the rest have either not posted a single status (9.20\%) or their instances were down at the time of crawl (11.58\%). In total, we gather 16,163,600 tweets, and 5,746,052 Mastodon statues.

\subsection{Followees}
\label{sec:followees}

We also crawl the user's followees for both Twitter and Mastodon accounts.
We use the Twitter Follows API\footnote{https://api.twitter.com/2/users/:id/following
} and the Mastodon Account's Following Endpoint\footnote{https://docs.joinmastodon.org/methods/accounts/\#following} respectively. 
Due to the rate limitations of the Twitter's API we crawl a sub-sample of 10\% of the migrated users.
For representativity, our sample relies on the followees distribution takes 5\% from above the median value and 5\% from below.
In total, we gather followee data for 13,068 users. This covers 11,453,484 followee relationships.

\subsection{Ethical Considerations}
\label{sec:ethicalconsiderations}

The datasets in this paper include both user and post information, and therefore might have privacy implications. To overcome any data mishandling, we have exclusively collected the publicly available data following well-established ethical procedures for social data. We have obtained a waiver from the ethics committee at the author’s institution.\footnote{anonymised for double-blind submission}
We anonymize the data before use and store it in a secure silo. Upon acceptance of the paper, anonymized data will be made available to the public, which we hope will help further works.

\section{RQ1: The Centralization Paradox}
\label{sec:instances}

\begin{figure}[t]
     \centering
     \includegraphics[width=0.8\columnwidth]{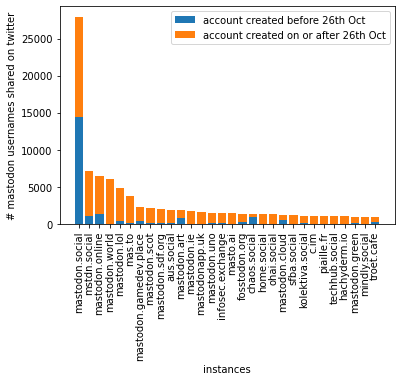}
     \caption{Top 30 Mastodon instances Twitter users migrated to.}
     \label{fig:accountsperinstance}
\end{figure}

\begin{figure}[t]
     \centering
     \includegraphics[width=0.8\columnwidth]{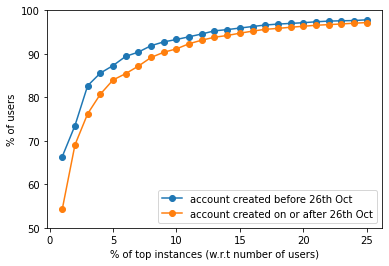}
     \caption{Percentage of users on top 25\% instances (w.r.t number of users).}
     \label{fig:accountsinbiginstances}
\end{figure}

Mastodon is a decentralized platform, in which users are (notionally) spread across thousands of independent instances. 
We therefore first test if the migration has resulted in true decentralization or if Mastodon experiences a paradox, whereby the majority of users centralize upon a small number of servers.
Overall, the Twitter users in our dataset migrate to 2,879 unique Mastodon instances. Figure~\ref{fig:accountsperinstance} presents a histogram of the number of users who have joined the top 30 Mastodon instances. 
The plot divide accounts into those created before the acquisition, and those created after.
Interestingly, not all the Mastodon accounts advertised on Twitter in response to Elon Musk's acquisition are new though. 21\% of the Mastodon accounts were created before the Musk's takeover.
Despite Mastodon's decentralization efforts,
we observe a clear trend towards centralization:
a large number of users migrate to a small set of instances. 
In particular, 
\url{mastodon.social}, a flagship Mastodon instance operated by Mastodon gGmbH, receives the largest fraction of migrated Twitter users.

We next explore the pressure towards Mastodon centralization by comparing the percentage of migrated users with the percentage of instances they join.
Figure~\ref{fig:accountsinbiginstances} plots the distribution of users across the top \% of instances. We find that nearly 96\% of users join the top 25\% of largest instances (w.r.t number of users). This centralization trend implies that a small number of instance owners, administrators and moderators have an impact on a large fraction of migrated users. 
Arguably, this means that Mastodon is at risk of becoming another (semi-)centralized service.

\begin{figure*}[t]
     \centering
     \includegraphics[width=0.9\textwidth]{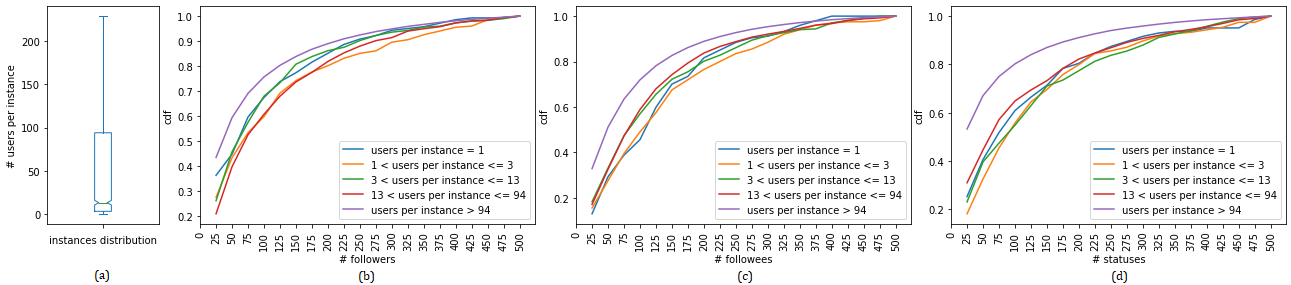}
     \caption{(a) Distribution of instances w.r.t to number of users. (b) CDF of number of followers of users on different-sized instances. (c) CDF of number of followees of users on different-sized instances. (d) CDF of number of statuses of users on different-sized instances.}
     \label{fig:socialnetworkwrtinstancesizeV1}
\end{figure*}

One possible explanation for this trend is that users intend to build a large social network by joining large and well-known instances. 
We therefore examine the relationship between the size of instances and social networks 
by  analyzing the number of followers and followees of users joining different-sized instances. 
We analyze all migrated users who join Mastodon after the Twitter acquisition and have 30 days old account (to ensure a fair comparison). This covers 50.59\% of all migrated users.
We divide the instances based on quantiles w.r.t number of users. Figure~\ref{fig:socialnetworkwrtinstancesizeV1} presents the distribution of instances by the number of users, CDFs of the number of followers, followees, and statuses of users on different-sized instances. 
Contrary to our hypothesis, users in the bigger instances tend to have smaller social networks. 
13.16\% of instances have just one user, who tends to have more followers, followees, and statuses than users in more populated instances.
Paradoxically,
the single user instances, 
have 64.88\% more followers,
follows 99.04\% more users,  and
posts 121.14\% more statuses (on average) than the users of the bigger instances. %
This implies that the size of an instance has a limited impact on the size of a user's social network. Rather it mainly depends on the user's activeness, engagement and networking. Hence, while large instances have more users, small instances attract more active users. 
Manual inspection suggests that this is because smaller instances have more dedicated and proactive user bases (whereas larger ones accumulate more experimental users).

\section{RQ 2: Social Network Influence}
\label{sec:socialnet}

There are at least two possible reasons for platform migration from Twitter to Mastodon, particularly after the Musk takeover:
\one A user might have decided to move for ideological reasons, if they disagree with Musk's actions after he gained control of Twitter;
and 
\two A user might have decided to move because a large fraction of the accounts they follow moved (and therefore Twitter has become irrelevant as a platform for them).
Of course, these two reasons are not contradictory or mutually exclusive. In this section, we attempt to distinguish between these reasons based on the observation that if a user moves because their immediate social network moves, a large proportion of their ego network neighbourhood would also have moved with them. We argue this offers an interesting example of social contagion.

\subsection{Twitter vs. Mastodon Social Network}
\label{sec:twittervsmastodonsocialnetwork}
\begin{figure}[t]
     \centering
     \includegraphics[width=0.8\columnwidth]{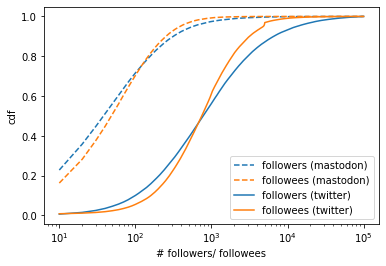}
     \caption{CDF of number of followers and followees of migrated users on Twitter and Mastodon.}
     \label{fig:twittervsmastodonsocialnetwork}
\end{figure}

We first analyze the size of the social network (\ie number of followers \& followees) that the migrated users have on both Twitter and Mastodon. 
Figure~\ref{fig:twittervsmastodonsocialnetwork} plots the CDF of the number of followers and followees of migrated users on both platforms. 
The median followers and followees that migrated users have on Twitter are 744 and 787, respectively.
Just 152 users (0.11\% of total migrated) have no Twitter followers, and 465 (0.35\% of total migrated) have no Twitter followees. 
In contrast, on Mastodon, 6.01\% of users have no followers, and 3.6\% do not follow anyone. The median followers and followees on Mastodon were 38 and 48, respectively. Interestingly, 1.65\% of migrated users gained a median of 33 \emph{more} followers on Mastodon than their Twitter followers. This confirms that these new users are yet to bootstrap a significant social network on Mastodon.
However, we emphasize that the median age of migrated accounts on Twitter is 11.5 years, in contrast to just 35 days on Mastodon. 
Hence, due to these disproportionate ages, the size of the social networks on the two platforms are not directly comparable.

\subsection{Social Network Driven Migration}
\label{sec:socialnetworkeffectinmigration}

\begin{figure}[t]
     \centering
     \includegraphics[width=0.8\columnwidth]{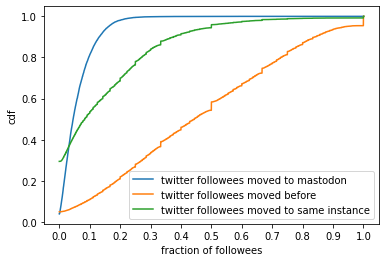}
     \caption{CDFs of the fraction of Twitter followees of each migrated user that (i) moved to Mastodon (blue) (ii) moved to Mastodon before the user (orange) and (iii) moved to the same instances on Mastodon as the user (green).}
     \label{fig:communitymigration}
\end{figure}

We next conjecture that a user's (Twitter) social network may have an impact on their likelihood of migration. For example, if a user's friends migrate to Mastodon, this may encourage them to do the same.

To inspect this, we analyze the followees data from both Twitter and Mastodon for 10\% of the migrated users (see \S{\ref{sec:followees}}). Figure~\ref{fig:communitymigration} shows CDFs of the fraction of Twitter followees of each migrated user that 
\one~moved to Mastodon (blue);
\two~moved to Mastodon before the user (orange);
and
\three~moved to the same Mastodon instances as the user (green). 
We notice that just 5.99\% of each user's followees also migrate (on average).
In fact, for 3.94\% of the migrated users, none of their Twitter followees move to Mastodon. 
Thus, the majority of the social network of the migrated users seems indeed reluctant to migrate, and sometimes they are the first in taking this step.

To better understand this, we compare the date on which each migrated user joined Mastodon with that of their Twitter followees who migrated as well.
We find that, out of their social network (\ie their followees),
4.98\% of the migrated users were the first
and 4.58\% were the last to migrate from Twitter to Mastodon. 
On average, 45.76\% of the followees of a user migrated to Mastodon before the user actually did.

We are also curious to understand if users select the same Mastodon instance as their social network. We therefore compare the instance of each migrated user with that of its Twitter followees.
On average, 14.72\% of each migrated user's followees (that move to Mastodon) join the same instance. 
With 15K+ Mastodon instances, this is a considerable proportion, suggesting a clear network effect. 
However, we also notice that this average is highly impacted by one flagship instance: \url{mastodon.social}. 
This is the largest instance available, and is probably the best known.
Of all the migrated users whose Twitter followees move to the same instance, 30.68\% are on \url{mastodon.social}. That said, we also find small instances that attract significant proportions of a given user's Twitter followers.
For example, 4.5\% of the migrated users whose Twitter followees join them on the same instance are on \url{mastodon.gamedev.place} (a Mastodon server focused on game development and related topics) %

\begin{figure}[t]
     \centering
     \includegraphics[width=0.8\columnwidth]{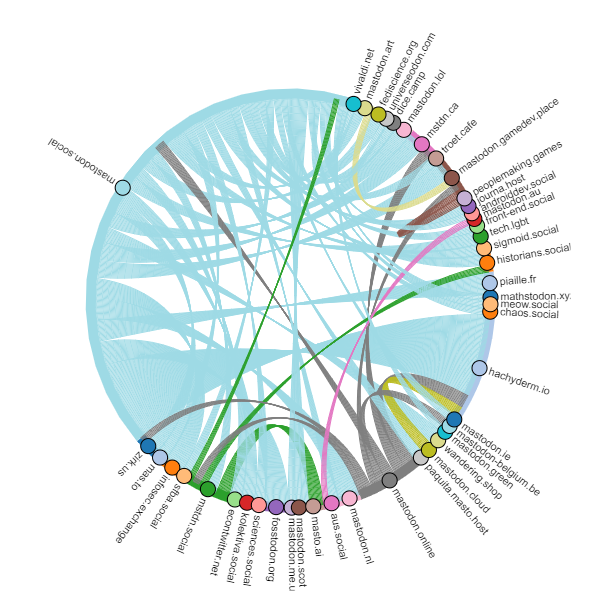}
     \caption{Chord plot of switching within Mastodon instances.}
     \label{fig:movedvisualizations}
\end{figure}

\begin{figure}[t]
     \centering
     \includegraphics[width=0.8\columnwidth]{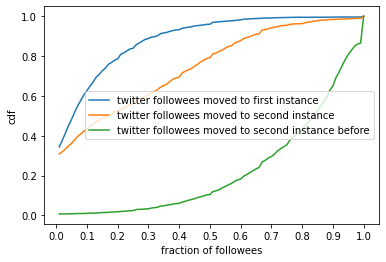}
     \caption{CDFs of the fraction of Twitter followees of each switched user that (i) moved to first instance (blue) (ii) moved to second instance (orange) and (iii) moved to second instance before the user (green).}
     \label{fig:switchingusersinfluence}
\end{figure}

\subsection{Instance Switching}

A unique feature of Mastodon is that users can easily `switch' instance. This involves migrating their data from one instance to another. 
We are curious to see if this is also driven by network effects.
Overall, 4.09\% of the users have switched from the Mastodon instance they initially created an account on (hereinafter first instance) to a new instance (hereinafter second instance).
Curiously, 97.22\% of these switches happened after Musk's Twitter takeover. 
This suggests that users may join initial instances, but migrate to a more suitable one once they are more experienced.

Figure~\ref{fig:movedvisualizations} shows the chord plot of switches from each user's first Mastodon instance to their second. A common pattern across these switches is that users move from general purpose/ flagship instances (\eg \url{mastodon.social}, \url{mastodon.online}) to more topic specific instances, \eg \url{sigmoid.social} (a Mastodon instance for people researching and working in Artificial Intelligence) and \url{historians.social} (a Mastodon server for people interested in history). 

Interestingly, we notice a strong social network influence behind these switches.
Figure~\ref{fig:switchingusersinfluence} shows the CDFs of the fraction of Twitter followees of each switched user that 
\one~moved to the first instance (blue);
\two~moved to the second instance (orange);
and 
\three~moved to second instance before the user (green). 
On average, 46.98\% of each user's followees (who moved to Mastodon) at some point also join the second instance. In contrast to just 11.4\% who join the first instance. 
Interestingly, 77.42\% of each switching user's followees (on average) joined the second instance before the user.
This suggests that the users switched from the first instance because a large fraction of their Twitter followees moved to the second one.

\section{RQ3: Timelines Analysis}
\label{sec:content}

We are next curious to understand how people use their (two) accounts after migration.

\subsection{Twitter vs. Mastodon Activity}
\label{sec:twittervsmastodonactivity}

\begin{figure}[t]
     \centering
     \includegraphics[width=0.8\columnwidth]{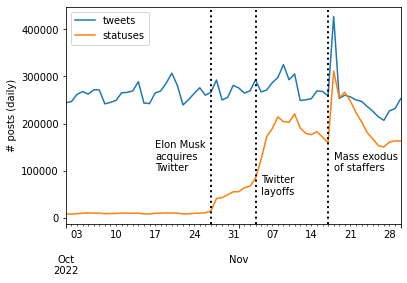}
     \caption{Temporal distribution of tweets and statuses posted by migrated users on Twitter and Mastodon respectively.}
     \label{fig:twittermastodonactivity}
\end{figure}

\begin{figure}[t]
     \centering
     \includegraphics[width=0.8\columnwidth]{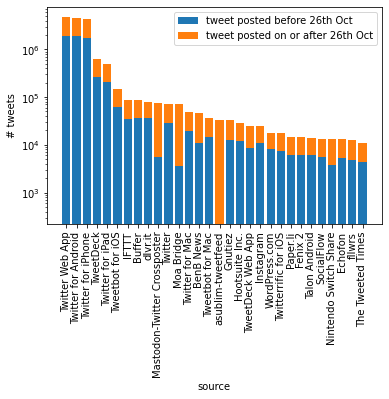}
     \caption{Top 30 sources of tweets. Note the log scale on the y-axis.}
     \label{fig:twitteractivitysource}
\end{figure}

\begin{figure}[t]
     \centering
     \includegraphics[width=0.8\columnwidth]{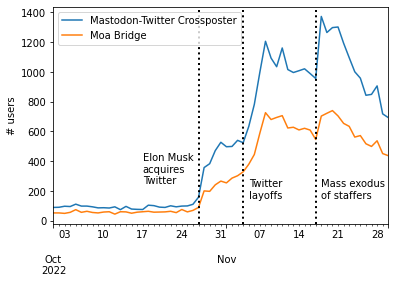}
     \caption{Number of users that use cross-posting tools daily.}
     \label{fig:twitteractivitysourceusers}
\end{figure}

We first analyze the timelines of migrated users from both Twitter and Mastodon. Figure~\ref{fig:twittermastodonactivity} shows the number of tweets on Twitter and the number of statuses on Mastodon posted by migrated users each day from October 01, 2022 to November 30, 2022. We observe a continuous growth in user activity on Mastodon after the acquisition of Twitter. 
However, the activity of migrated users on Twitter do not decrease in parallel, \ie our migrated users are using both their Twitter and Mastodon accounts simultaneously.

\begin{figure}[t]
     \centering
     \includegraphics[width=0.8\columnwidth]{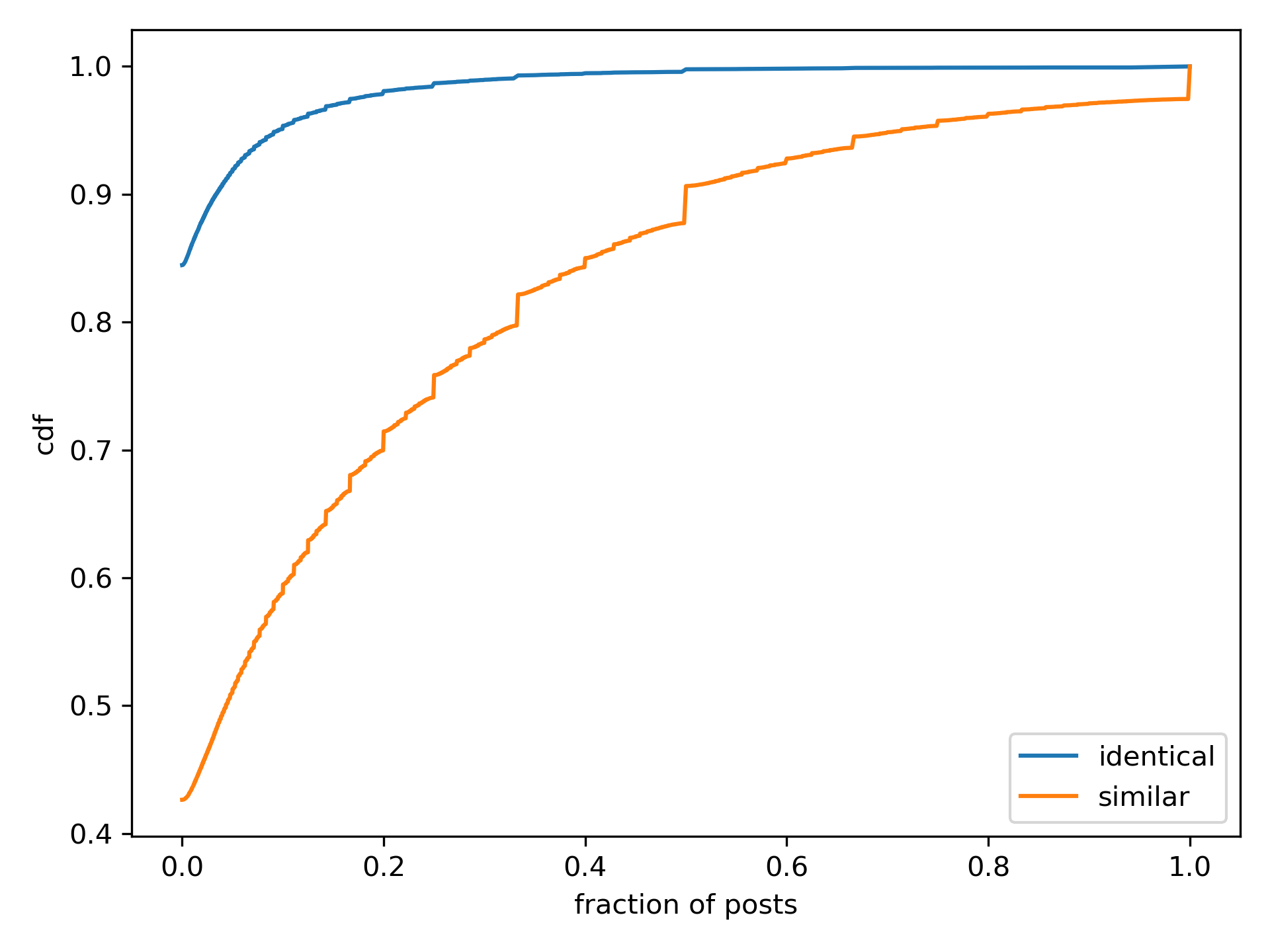}
     \caption{CDFs of fraction of each migrated user's Mastodon statuses that are identical or similar to its tweets.}
     \label{fig:identicalsimilar}
\end{figure}

We next check if people are generating identical content across both platforms or are, instead, projecting multiple `personas'.
Figure~\ref{fig:identicalsimilar} plots the CDFs of the fraction of each migrated user's Mastodon statuses that are identical or similar to its tweets. 
We consider the Mastodon status similar to a tweet if the cosine-similarity of their sentence embeddings~\cite{reimers-2019-sentence-bert} is greater than 0.7. 
Surprisingly, just 1.53\% of each migrated user's Mastodon statuses are identical.
On average, just 16.57\% of each user's Mastodon statues are similar to their tweets.
Instead, 84.45\% of the migrated users use the two platforms to post completely different content. 
This suggests a mix of users, some of whom create different personas on the two platforms, and a smaller subset who mirror all their content.

A potential explanation for the latter is the use of cross-posting tools. 
Such tools allow users to automatically mirror their Mastodon statues on Twitter, and vice versa.
To examine this, we compare the number of tweets posted via different sources before and after Musk's takeover in Figure~\ref{fig:twitteractivitysource}.
Naturally, the majority are posted by official Twitter clients such as the Twitter Web App. The two sources that increase most dramatically, however, are two well-known cross-posters, Mastodon-Twitter Crossposter and Moa Bridge --- by 1128.95\% and 1732.26\%, respectively.
Of all migrated users, 5.73\% use one of the two cross-posters at least once.
This suggested such users see both Twitter and Mastodon as viable platforms, and have limited intention of creating multiple 'personas'.
Figure~\ref{fig:twitteractivitysourceusers} also plots the number of users using cross-posters over time. 
We see that their usage increases rapidly after Musk's takeover. The downward trend towards the end of November is likely a result of the posting issues that cross-posters faced after their posting rate limit was revoked by Twitter~\cite{crossposterexit}.

\subsection{Hashtags}

\begin{figure}[t]
     \centering
     \includegraphics[width=0.8\columnwidth]{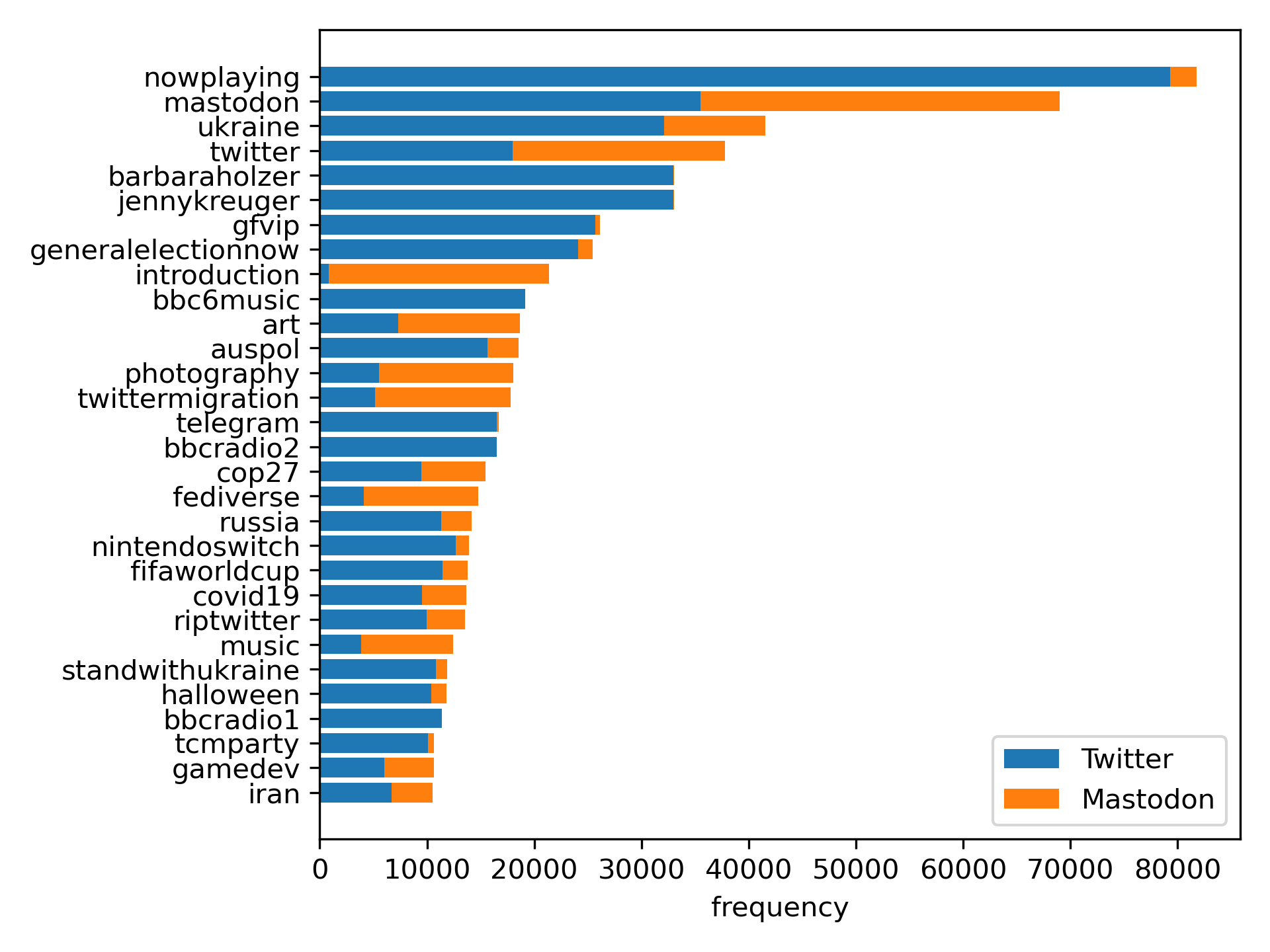}
     \caption{Top 30 hashtags along with their frequencies on Twitter and Mastodon.}
     \label{fig:top30hashtags}
\end{figure}

Given that 84.45\% of the migrated users post completely different content on the two platforms, we next inspect the hashtags used.
This gives us a flavour of the parallel discussions taking place on Mastodon and Twitter.
Figure~\ref{fig:top30hashtags} presents the top 30 most frequent hashtags used over the two platforms by the migrated users. We notice that users discuss more diverse topics on Twitter such as Entertainment (\#NowPlaying, \#BBC6Music), Celebrities (\#BarbaraHolzer), and Politics (\#StandWithUkraine, \#GeneralElectionNow), whereas  Mastodon seems dominated by Fediverse related discussion (\#fediverse) and the migration to it (\#TwitterMigration). 
We conjecture that we might see more diverse discussions on Mastodon once the migrated users make themselves familiar with the platform.

\begin{figure}[t]
     \centering
     \includegraphics[width=0.8\columnwidth]{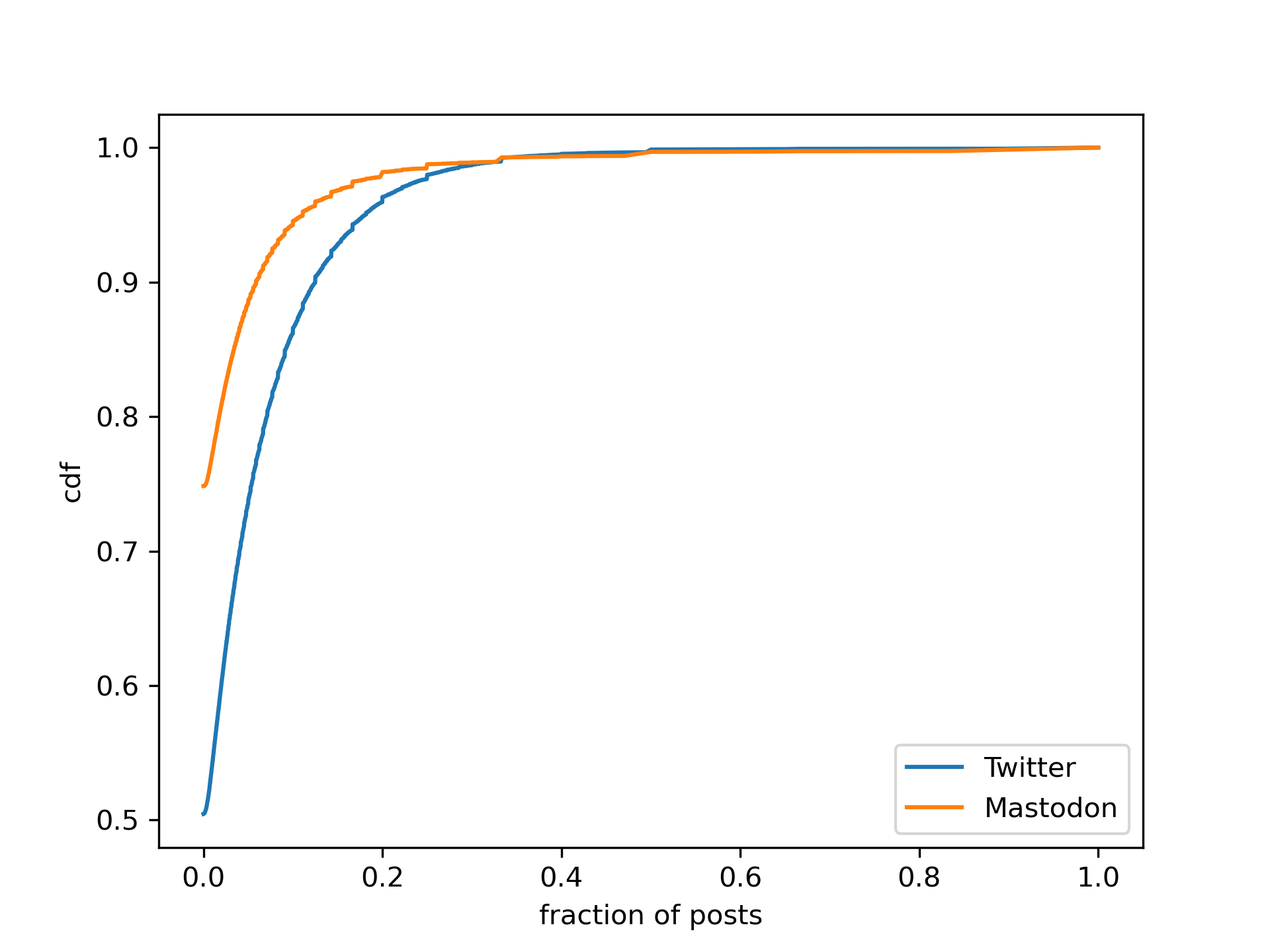}
     \caption{CDFs of fraction of each migrated user's toxic posts on Twitter and Mastodon.}
     \label{fig:toxicity}
\end{figure}

\subsection{Toxicity Analysis}

Moderation on Mastodon has received significant attention in recent months~\cite{bin2022toxicity,anaobi2023will}. 
This is because the administrators of Mastodon instances do not universally have the resources to moderate malicius content.
To shed light on this, we study the extent to which toxic content is  shared by migrated users on both platforms. To do this, we label all tweets and statuses using Google Jigsaw’s Perspective API.\footnote{https://www.perspectiveapi.com} 
For a given post, Perspective returns a score between 0 and 1 for its toxicity (0 $=$ non-toxic). Specifically, we use the API's TOXICITY attribute that defines toxicity as ``a rude, disrespectful, or unreasonable comment that is likely to make people leave a discussion''. In the literature, 0.5 is the most common choice to threshold the perspective scores~\cite{bin2022toxicity,rottger2020hatecheck,papasavva2020raiders}, however, higher values such as 0.8 are also used~\cite{agarwal2021hate}. Here, we use 0.5 as a threshold and consider a post to be toxic if its toxicity score is greater than 0.5 (and vice versa).

Figure~\ref{fig:toxicity} shows the CDFs of the fraction of each migrated user's toxic posts on Twitter and Mastodon. Overall, just 5.49\% of tweets are toxic.
Mastodon is substantially less toxic, with just 2.80\%. On average, each user posts 4.02\% toxic tweets on Twitter vs. just 2.07\% toxic statuses on Mastodon. Even though the discourse is non-toxic over both platforms, we notice that 14.26\% of migrated users post at least one toxic post on both the platforms. While this may not be problematic for Twitter which has its own moderation team, it might present challenges for Mastodon, where volunteer administrators are responsible for content moderation~\cite{hassan2021exploring}.

\section{Related Work}
\label{sec:relatedwork}

\pb{Decentralised Social Networks.} 
Many previous efforts have been made to build decentralized online social platforms. In the earliest days, there were many peer-to-peer online social networks, such as Safebook \cite{cutillo2009safebook}, PeerSoN \cite{buchegger2009peerson}, LotusNet \cite{aiello2012lotusnet}, and LifeSocial.KOM \cite{graffi2011lifesocial}. 
However, performance and security limitations~\cite{paul2014survey},
limited their adoption and success. 
New decentralized social networks, such as Mastodon, Pleroma, Pixelfed, and PeerTube, have since emerged.
In sum, these platforms are referred to as the \emph{Fediverse}. 
These social network applications use ActivityPub, a W3C protocol, to implement server federation. 
Some recent work has looked into these new decentralized social networks.
For instance, a large-scale measurement study of Mastodon~\cite{raman2019challenges}, found centralization trends in Mastodon. Paradoxically, we found that while centralization occurs in terms of how many users are attracted to an instance, smaller instances attract more active users.
Other works focus on user behavior across instances~\cite{la2022network,la2022information}.
Our work also touches upon the need for decentralised moderation.
This has been investigated in prior work on Pleroma (another Fediverse microblog.
Hassan et al identify novel challenges~\cite{hassan2021exploring} and propose a strawman solution.
Zia et al.~\cite{bin2022toxicity} also propose model sharing solution to help automate moderation. 
Our work confirms the presence of toxic content in Mastodon, though the numbers identified do not show a trend towards greater toxicity than Twitter.

\pb{Social Network Migration.} There have been a number of measurement studies on social network migration. For example, \cite{fiesler2020moving} measured the migration activity of fandom, tracking migrating users and the reasons behind their migration. The authors find that policy and value-based aspects are determinant in the migration.
Gerhart et al.~\cite{gerhart2019social} analyze user migration from traditional social networks to anonymous social networks perspective.
They identify that social norms drive migration.
Otala et al.~\cite{m2021political} study the migration of Twitter users to Parler. 
The results show that, although Parler is not widely used, it has a significant impact on political polarization.
Our work also studies the migration of Twitter users.
However, to the best of our knowledge, it is the first to systematically measure and analyze the migration of users from centralised Twitter to a decentralised platform.

\section{Conclusion}
\label{sec:conslusion}

In this paper, we have explored the migration of users from Twitter to Mastodon, prompted by Elon Musk's acquisition of Twitter. 
We have focused on three RQs:   \one~How are new users spread across Mastodon instances, and are there any consequences for decentralization?
\two~How much (if at all) does a user's ego-centric Twitter network influence their migration to Mastodon?
\three~What are usage patterns of migrated users across both platforms?
To answer \textbf{RQ1}, we have found that 2.26\% of users completely left Twitter, deleting their account. Despite Mastodon's decentralized architecture, we found that the 25\% largest instance on Mastodon contains 96\% of the users. Paradoxically, while larger instances attract more users, smaller ones attract more active users, reinforcing Mastodon's decentralization.
To answer \textbf{RQ2}, we showed that the size of the Mastodon instance had limited effect on the size of the user's social network. 
We observed the impact of social network in migration, with an average of 14.72\% of Twitter followees per user migrating to the exact same Mastodon instance as user.
To answer \textbf{RQ3}, we found that users tend to post \emph{different} content across platforms. On average, only 1.53\% of Mastodon posts per user were identical to Twitter. In terms of toxicity, most of the user's content on both platforms was non-toxic. Mastodon appears to be less toxic than Twitter though.  Overall, just 5.49\% of tweets and 2.80\% of statuses posted by migrated users on Twitter and Mastodon respectively were toxic.

There are a number of lines of future works. We would like to further investigate whether migrating users retain their Mastodon accounts or return to Twitter, and whether new users are joining the migration wave. It will be interesting to see what the future holds for these user-driven centralized Mastodon instances. This study provides the first step in the migration of Twitter to Mastodon. We hope that it will inspire further exploration of follow-up work.

\bibliographystyle{abbrv}
\bibliography{flocking-mastodon.bib}
\end{document}